\documentclass[conference]{IEEEtran}
\IEEEoverridecommandlockouts

\usepackage{cite}
\usepackage{amsmath,amssymb,amsfonts}
\usepackage{algorithmic}
\usepackage{graphicx}
\usepackage{textcomp}
\usepackage{xcolor}
\usepackage{booktabs}
\usepackage{array}
\usepackage{tikz}
\usetikzlibrary{shapes.geometric, arrows.meta, positioning, calc, fit, backgrounds}

\def\BibTeX{{\rm B\kern-.05em{\sc i\kern-.025em b}\kern-.08em
    T\kern-.1667em\lower.7ex\hbox{E}\kern-.125emX}}

\tikzset{
  block/.style   = {rectangle, rounded corners=2pt, draw=black!70, fill=blue!8,
                    text width=6.0em, minimum height=2.0em, align=center,
                    font=\scriptsize, inner sep=3pt},
  agent/.style   = {rectangle, rounded corners=2pt, draw=black!70, fill=orange!12,
                    text width=6.4em, minimum height=2.0em, align=center,
                    font=\scriptsize, inner sep=3pt},
  decision/.style= {diamond, aspect=2, draw=black!70, fill=green!10,
                    text width=4.2em, align=center, font=\scriptsize, inner sep=1pt},
  io/.style      = {rectangle, draw=black!70, fill=gray!10,
                    text width=6.0em, minimum height=1.8em, align=center,
                    font=\scriptsize, inner sep=3pt},
  arr/.style     = {-{Stealth[length=4pt]}, thick, draw=black!75},
  fb/.style      = {-{Stealth[length=4pt]}, thick, draw=red!65, dashed}
}

\begin{document}

\title{Agent-Orchestrated Adaptive RAG: A Comparative\\ Study on Structured and Multi-Hop Retrieval}

\author{
\IEEEauthorblockN{Anuj Maharjan}
\IEEEauthorblockA{\textit{Dept. of Electrical Engineering}\\\textit{and Computer Science}\\
\textit{University of Toledo}\\
Toledo, OH, USA \\
anuj.maharjan@utoledo.edu}
\and
\IEEEauthorblockN{Devinder Kaur}
\IEEEauthorblockA{\textit{Dept. of Electrical Engineering}\\\textit{and Computer Science}\\
\textit{University of Toledo}\\
Toledo, OH, USA \\
devinder.kaur@utoledo.edu}
\and
\IEEEauthorblockN{Richard G. Molyet}
\IEEEauthorblockA{\textit{Dept. of Electrical Engineering}\\\textit{and Computer Science}\\
\textit{University of Toledo}\\
Toledo, OH, USA \\
richard.molyet@utoledo.edu}
}

\maketitle

\begin{abstract}
Retrieval-Augmented Generation (RAG) enhances Large Language Models (LLMs) by grounding their responses in external knowledge, but conventional pipelines rely on static, single-step retrieval that limits performance on complex queries. This paper presents an Agent-Orchestrated Adaptive RAG framework that introduces dynamic query decomposition, iterative retrieval, and a bounded self-reflective evaluation loop. We evaluate the system across two complementary datasets: a domain-specific DevOps knowledge base and the multi-hop reasoning benchmark MuSiQue. Using metrics that include overall score, citation accuracy, mean reciprocal rank, and topic coverage, we find that query decomposition yields consistent gains in the structured domain (overall score $+0.04$, MRR $+0.17$ on DevOps) but degrades ranking precision on the multi-hop benchmark, while the reflection mechanism improves citation accuracy at a substantial latency cost. These contrasting results show that agentic enhancements are not universally beneficial and must be applied selectively according to query and domain characteristics. Our findings argue for adaptive, cost-aware orchestration rather than uniformly aggressive reasoning pipelines.
\end{abstract}


% =====================================================
\section{Introduction}
% =====================================================
Large Language Models (LLMs) have shown strong capabilities in natural language understanding and generation, yet they remain constrained by two persistent limitations: factual hallucination and an inability to access knowledge beyond their training data \cite{ji2023hallucination, mialon2023augmented}. These limitations are especially consequential in high-stakes operational settings such as DevOps and Site Reliability Engineering (SRE), where the correct interpretation of technical documentation can be the difference between rapid incident resolution and prolonged downtime \cite{amazon2024chatops}.

Retrieval-Augmented Generation (RAG) addresses these limitations by allowing a model to consult an external knowledge base before generating a response \cite{lewis2020rag, guu2020realm}. Conventional \emph{Naive} RAG follows a fixed, single-pass procedure: the query is embedded, the top-$k$ most similar chunks are retrieved, and these are prepended to the prompt as context \cite{amorim2024empirical}. This works well for isolated factual lookups, but it struggles with queries that require reasoning across several documents or chaining intermediate results \cite{feldman2023multihop, gao2023survey}.

Agentic RAG extends this paradigm by delegating retrieval to autonomous agents that can plan, retrieve iteratively, and critique their own output \cite{ghadekar2025survey, wang2024evolution}. Instead of a single retrieve-then-generate step, an agentic system can decompose a complex query into sub-questions, gather evidence for each, and refine its answer through self-reflection \cite{asai2024selfrag, yao2022react}. Interest in this direction has grown quickly, but most studies evaluate agentic RAG on a single domain or benchmark. As a result, it remains unclear how these enhancements behave across domains that differ in structure and reasoning demand.

This paper addresses that gap through a deliberately dual-dataset evaluation. We make the following contributions:

\begin{enumerate}
  \item We implement a complete Agent-Orchestrated Adaptive RAG system composed of a Query Classifier, a Query Decomposer, an Answer Evaluator, and a central Orchestrator, built on a fully local, privacy-first inference stack \cite{meta2024llama3, gerganov2023gguf}.
  \item We evaluate the system on two complementary datasets: a custom DevOps knowledge base that represents structured, domain-specific retrieval, and MuSiQue \cite{trivedi2022musique}, a compositional multi-hop benchmark. This pairing lets us observe agentic behavior under contrasting conditions.
  \item We characterize the tradeoffs introduced by decomposition and reflection, showing empirically that decomposition is domain-dependent and that reflection adds significant latency for inconsistent quality gains.
\end{enumerate}

% =====================================================
\section{Background and Related Work}
% =====================================================

\subsection{Generative Models and LLMs}
Generative models learn an underlying data distribution and sample novel outputs from it, a lineage that runs from generative adversarial networks \cite{goodfellow2014gan, gui2021gansurvey} to the autoregressive language models that dominate today \cite{brown2020gpt3}. A language model factorizes the probability of a token sequence and predicts each token conditioned on its predecessors,
\begin{equation}
P(x_n \mid x_1, x_2, \ldots, x_{n-1}),
\end{equation}
which lets it generate coherent, contextually appropriate text across domains.

The success of modern LLMs is largely attributable to the Transformer architecture \cite{vaswani2017attention}, which replaced recurrence with attention. The self-attention operation weighs the relevance of every token to every other token,
\begin{equation}
\mathrm{Attention}(Q,K,V) = \mathrm{softmax}\!\left(\frac{QK^{\top}}{\sqrt{d_k}}\right) V,
\end{equation}
where $Q$, $K$, and $V$ are the query, key, and value projections and $d_k$ is the key dimensionality. Multi-head attention runs several such operations in parallel to capture diverse linguistic relationships, while feedforward sublayers and positional encodings \cite{shaw2018relative, dufter2022position} supply non-linearity and sequence order. Input text is first segmented by subword tokenization such as byte-pair encoding \cite{sennrich2015bpe} and embedded into dense vectors before passing through the stacked Transformer layers \cite{devlin2018bert, radford2018gpt}. At generation time, decoding strategies---greedy selection, beam search \cite{graves2012sequence}, and stochastic methods such as temperature scaling and nucleus sampling \cite{holtzman2019degeneration}---balance coherence against diversity.

Despite their fluency, these models exhibit well-documented limitations: they hallucinate and lack grounding in verifiable sources \cite{ji2023hallucination}, can encode social and cultural bias from their training data \cite{bender2021parrots}, and are costly to train and serve \cite{strubell2019energy}. These limitations are precisely what retrieval-based augmentation is designed to mitigate.

\subsection{Naive and Advanced RAG}
The foundational RAG framework \cite{lewis2020rag} couples the parametric knowledge stored in an LLM's weights with non-parametric knowledge retrieved from an external corpus, and demonstrated that grounding generation in retrieved evidence reduces hallucination and improves factual accuracy. A standard pipeline has three components. The \emph{retriever} fetches relevant documents, typically via dense retrieval in which queries and documents are encoded by neural encoders \cite{mikolov2013distributed, mikolov2013word2vec, reimers2019sbert} and compared in vector space, as in Dense Passage Retrieval \cite{karpukhin2020dpr} with similarity $\mathrm{sim}(q,d) = q^{\top} d$. The \emph{augmentation} step injects the retrieved passages into the prompt, and the \emph{generator}---the LLM---produces a response conditioned on both the query and the retrieved context.

Naive RAG retrieves a fixed top-$k$ set and appends it directly, which is simple but admits irrelevant or redundant passages \cite{amorim2024empirical}. Advanced RAG counters this with query rewriting and expansion \cite{wang2023query2doc}, re-ranking of candidates, and chunking strategies that improve context selection \cite{gao2023survey}. Recent empirical work on chunking, retrieval, and re-ranking for policy-document question answering confirms that these pipeline design choices substantially shape downstream answer quality \cite{maharjan2026chunking}. Because exhaustive nearest-neighbor search becomes prohibitive as a corpus grows, efficient retrieval relies on approximate nearest-neighbor indexes such as FAISS \cite{johnson2019billion, douze2024faiss, ghadekar2023faiss}, which trade a small recall loss for large gains in latency. Multi-hop RAG variants extend this further by retrieving iteratively rather than once, refining the context across reasoning steps \cite{feldman2023multihop}.

\subsection{Agentic RAG}
Surveys have charted the shift from static pipelines to agentic systems capable of dynamic tool use, iterative retrieval, and self-correction \cite{ghadekar2025survey, wang2024evolution, mishra2026sok}. Whereas a Naive pipeline executes a predetermined sequence, an agentic system determines its control flow at runtime, using intermediate outputs to guide subsequent actions and invoking external tools only as needed \cite{wang2023agents}. Agentic RAG systems are typically organized around four interacting capabilities: \emph{query understanding and planning}, which interprets the request and selects a solving strategy; \emph{tool and retrieval orchestration}, which decides when and how to retrieve; \emph{multi-hop reasoning and decomposition}, which breaks complex problems into sub-tasks whose results are later combined; and \emph{self-reflection}, which critiques generated output and triggers refinement when issues are detected \cite{ng2024agentic, yan2024multiagent}.

These capabilities have concrete instantiations in the literature. The ReAct framework \cite{yao2022react} formalized the interleaving of reasoning traces and actions, Toolformer \cite{schick2023toolformer} showed that models can learn to call external tools, and Self-RAG \cite{asai2024selfrag} introduced learned reflection signals for selective retrieval and critique. Corrective RAG \cite{yan2024corrective} added an explicit retrieval-correction step, and Adaptive-RAG \cite{jeong2024adaptive} routed queries to retrieval strategies of differing complexity based on predicted question difficulty \cite{wandb2024agentic, hu2024selfcorrection}. Our system shares this lineage but is distinguished by an explicit multi-agent decomposition architecture that we evaluate across two contrasting domains rather than a single benchmark.

\subsection{Reasoning and Multi-Hop Retrieval}
Several prompting frameworks elicit explicit reasoning in LLMs. Chain-of-Thought prompting \cite{wei2022cot} encourages the model to generate intermediate reasoning steps before committing to an answer, improving performance on tasks that require composition. Least-to-Most prompting \cite{zhou2023leasttomost} decomposes a hard problem into a sequence of simpler sub-problems solved in order, an approach that aligns closely with the query decomposition strategy adopted in this work. Self-refinement methods iterate on a model's own draft using its self-generated feedback \cite{madaan2023selfrefine}, and inference-time self-correction has been surveyed as a general mechanism for improving reliability \cite{hu2024selfcorrection}.

Multi-hop question answering benchmarks provide the testbed for these methods. HotpotQA \cite{yang2018hotpotqa} popularized explainable multi-hop evaluation, and MuSiQue \cite{trivedi2022musique} was deliberately constructed by composing single-hop questions so that answering requires genuine evidence integration across independent documents and resists single-hop shortcuts. Tree- and graph-structured retrieval has been proposed to better capture the dependencies among reasoning steps \cite{shi2026reasoning}. Despite this activity, comparatively few studies place a multi-hop benchmark alongside a structured, domain-specific corpus to isolate how the same agentic mechanisms behave under different reasoning demands---the contrast this work is designed to expose.

% =====================================================
\section{System Architecture}
% =====================================================
The proposed system extends the conventional RAG pipeline with an agent-orchestrated control layer capable of adaptive routing, query decomposition, and bounded self-correction. Rather than committing to a single retrieval strategy, the system selects a strategy per query through coordinated interaction among specialized agents.

\subsection{Overview}
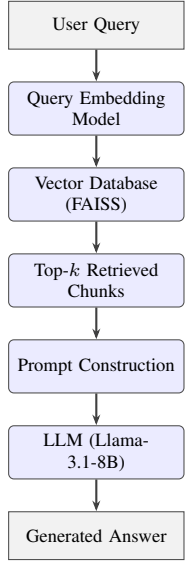
\begin{figure}[t]
\centering
\begin{tikzpicture}[node distance=4.2mm and 0mm]
  \node[io]    (q)   {User Query};
  \node[block] (emb) [below=of q]   {Query Embedding Model};
  \node[block] (db)  [below=of emb] {Vector Database (FAISS)};
  \node[block] (topk)[below=of db]  {Top-$k$ Retrieved Chunks};
  \node[block] (pc)  [below=of topk]{Prompt Construction};
  \node[block] (llm) [below=of pc]  {LLM (Llama-3.1-8B)};
  \node[io]    (ans) [below=of llm] {Generated Answer};
  \draw[arr] (q)--(emb); \draw[arr] (emb)--(db); \draw[arr] (db)--(topk);
  \draw[arr] (topk)--(pc); \draw[arr] (pc)--(llm); \draw[arr] (llm)--(ans);
\end{tikzpicture}
\caption{Standard Naive RAG pipeline: a single-pass retrieval process with no adaptive control or iterative reasoning.}
\label{fig:naive}
\end{figure}

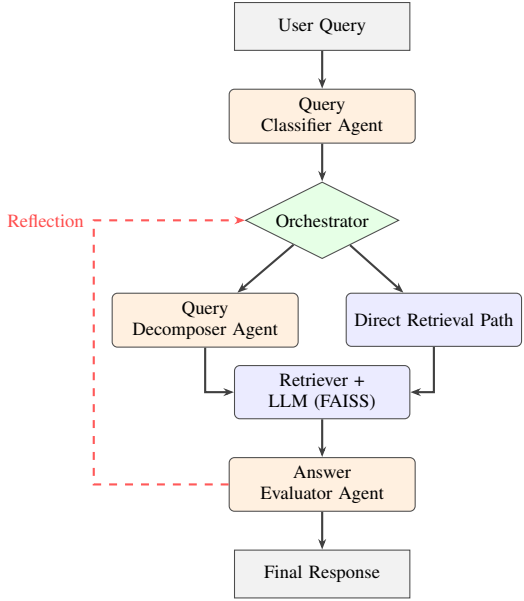
\begin{figure}[t]
\centering
\begin{tikzpicture}[node distance=5mm and 7mm]
  \node[io]       (q)    {User Query};
  \node[agent]    (cls)  [below=of q] {Query Classifier Agent};
  \node[decision] (orc)  [below=of cls] {Orchestrator};
  \node[agent]    (dec)  [below left=7mm and -2mm of orc] {Query Decomposer Agent};
  \node[block]    (dir)  [below right=7mm and -2mm of orc] {Direct Retrieval Path};
  \node[block]    (ret)  [below=14mm of orc] {Retriever + LLM (FAISS)};
  \node[agent]    (eval) [below=of ret] {Answer Evaluator Agent};
  \node[io]       (fin)  [below=of eval] {Final Response};

  \draw[arr] (q)--(cls);
  \draw[arr] (cls)--(orc);
  \draw[arr] (orc)-- (dec);
  \draw[arr] (orc)-- (dir);
  \draw[arr] (dec)|-(ret);
  \draw[arr] (dir)|-(ret);
  \draw[arr] (ret)--(eval);
  \draw[arr] (eval)--(fin);
  % reflection feedback loop
  \draw[fb] (eval.west) -| ($(orc.west)+(-2,0)$) |- (orc.west)
    node[midway, left, font=\scriptsize, text=red!70] {Reflection};
\end{tikzpicture}
\caption{Proposed Agent-Orchestrated Adaptive RAG architecture, showing dynamic routing, decomposition, and a bounded self-reflective evaluation loop (dashed feedback).}
\label{fig:agentic}
\end{figure}

Figure~\ref{fig:naive} shows the Naive RAG baseline for reference, and Fig.~\ref{fig:agentic} shows the proposed architecture. A central Orchestrator routes each query, invoking only the agents a given query requires. Two adaptive behaviors sit on top of the baseline retrieval path: decomposition-based reasoning for multi-hop queries and reflection-based correction for quality assurance.

% ---------- Figure 1: Naive RAG (TikZ) ----------

% ---------- Figure 2: Agentic architecture (TikZ) ----------

\subsection{Orchestrator and Control Logic}
The Orchestrator is the central controller. It routes queries based on the classifier output, triggers decomposition when required, and manages the evaluation--reflection cycle. Its decision logic is rule-based:

\begin{enumerate}
  \item If the query is simple $\rightarrow$ direct retrieval.
  \item If the query is complex $\rightarrow$ decomposition pipeline.
  \item If evaluation fails $\rightarrow$ trigger reflection (at most two retries).
\end{enumerate}

This design ensures that the system incurs additional computation only when a query warrants it, preserving efficiency for the common simple case.

\subsection{Query Classification and Metadata Filtering}
The Query Classifier Agent categorizes each incoming request into document types, enabling metadata-aware filtering before similarity ranking \cite{wandb2024agentic}. Because the vector store keeps document metadata alongside dense embeddings, filtering by attributes such as document type (for example, restricting retrieval to runbooks or incident reports) narrows the candidate space and improves semantic precision. This separates Agentic RAG from purely similarity-driven Naive pipelines, which cannot reason over structured attributes.

\subsection{Query Decomposition}
The Query Decomposer Agent enables multi-hop reasoning by splitting a complex query into ordered sub-queries \cite{zhou2023leasttomost, shi2026reasoning}. The workflow is: (i) detect multi-hop structure, flagging queries that involve multiple entities or chained dependencies; (ii) generate ordered sub-queries such that earlier results can inform later retrieval; (iii) retrieve independently for each sub-query; and (iv) aggregate the intermediate answers into a single coherent response that aligns with the original intent.

\subsection{Answer Evaluation and Reflection}
The Answer Evaluator Agent adds a self-reflective layer that critiques generated responses along four axes: relevance to the query, citation accuracy, support (grounding in retrieved context), and hallucination detection \cite{asai2024selfrag, madaan2023selfrefine}. When the evaluator flags an issue, the Orchestrator reprocesses the query---refining the prompt or adjusting the retrieved context by selecting different chunks or re-ranking results. To bound cost and latency, the reflection loop is capped at two iterations, balancing potential quality gains against responsiveness \cite{hu2024selfcorrection}.

\subsection{Data Preprocessing and Indexing}
The quality of a RAG system depends as much on how its corpus is prepared as on the model that consumes it; retrieval failures are frequently attributable to preprocessing and segmentation rather than to the generator \cite{lewis2020rag, gao2023survey}. Heterogeneous corpora are particularly challenging, since standard text extraction often discards structural cues---table headers, hierarchical lists, contextual groupings---that are essential for accurate retrieval. We therefore convert source files (PDF, DOCX, YAML) to structured Markdown using IBM Docling \cite{ibm2025docling, ibm2025doclingtech}, which uses layout-aware models to preserve visual hierarchy and decompose documents into semantic blocks rather than emitting an undifferentiated stream of text.

A custom preprocessing layer then enriches each document with YAML frontmatter encoding attributes such as document type and service, which the vector store later uses for attribute-based filtering \cite{wandb2024agentic}. This metadata is what allows the agentic pipeline to reason over structure, not just surface similarity. Documents are segmented with a token-based semantic chunking strategy: 600-token chunks with a 100-token overlap. The overlap preserves contextual continuity across boundaries, and the chunk size was selected empirically to balance three competing constraints---the LLM context window, the retrieval discrimination capacity of the index, and the preservation of technical coherence. Segmentation involves a recall--precision tradeoff: larger chunks improve semantic completeness while smaller chunks sharpen retrieval specificity, and improper segmentation can isolate evidence needed to reconstruct a multi-hop reasoning chain.

\begin{figure}[t]
\centering
\begin{tikzpicture}[node distance=3.6mm and 0mm]
  \node[io]    (raw) {Raw Documents (PDF, DOCX, YAML)};
  \node[block] (doc) [below=of raw] {Docling Structural Conversion};
  \node[block] (md)  [below=of doc] {Structured Markdown Output};
  \node[block] (yaml)[below=of md]  {YAML Frontmatter Injection};
  \node[block] (chk) [below=of yaml]{Semantic Chunking (600 / 100 overlap)};
  \node[block] (emb) [below=of chk] {Embedding Generation (BGE-base-en-v1.5)};
  \node[block] (idx) [below=of emb] {FAISS Index Construction};
  \draw[arr] (raw)--(doc); \draw[arr] (doc)--(md); \draw[arr] (md)--(yaml);
  \draw[arr] (yaml)--(chk); \draw[arr] (chk)--(emb); \draw[arr] (emb)--(idx);
\end{tikzpicture}
\caption{Document ingestion and preprocessing pipeline for the proposed DevOps knowledge base.}
\label{fig:ingest}
\end{figure}
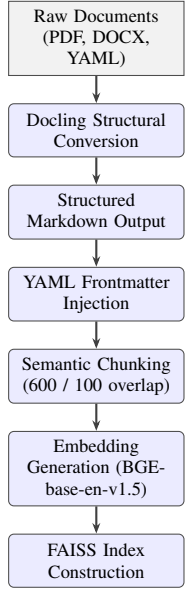

Each chunk is embedded with the BGE encoder and indexed in FAISS together with its metadata, enabling hybrid filtering in which attribute constraints (for example, restricting to runbooks or incident reports) are applied before similarity ranking to narrow the candidate space. In the agentic setting the index is not a passive store but an interactive component: agents may issue multiple retrieval calls, adjust filters, and re-rank results across reasoning steps, turning retrieval from a single lookup into a sequential decision process. The full ingestion pipeline---structural conversion, frontmatter injection, chunking, embedding, and index construction---is summarized in Fig.~\ref{fig:ingest}.

% ---------- Figure: ingestion pipeline (TikZ) ----------

\subsection{Implementation and Knowledge Base}
The generative core is Meta's Llama-3.1-8B-Instruct, quantized to 4-bit GGUF for local inference via \texttt{llama.cpp} \cite{meta2024llama3, gerganov2023gguf}. Retrieval uses the BAAI \texttt{BGE-base-en-v1.5} embedding model \cite{baai2023bge, reimers2019sbert}, chosen for its strong performance on semantic-similarity benchmarks, and vector search is handled by FAISS \cite{johnson2019billion}. The entire stack runs locally, a deliberate privacy-first choice for the DevOps setting, where operational documentation is often sensitive. Similarity between a query embedding $q$ and a document embedding $d_i$ is computed by cosine distance,
\begin{equation}
\mathrm{sim}(q, d_i) = \frac{q \cdot d_i}{\lVert q \rVert\, \lVert d_i \rVert},
\end{equation}
which is scale-invariant and well-suited to nearest-neighbor search.

The DevOps knowledge base comprises 80 documents (approximately 10{,}000 words) spanning six documentation types---Standard, Architecture, Runbooks, Incidents, Postmortem, and Onboarding---generated with LLM assistance and validated manually to reflect realistic organizational knowledge. Although domain-specific, the corpus is constructed to emulate broader properties of real knowledge bases, including structural heterogeneity, implicit cross-document relationships, and multi-step reasoning requirements.

% =====================================================
\section{Experimental Setup}
% =====================================================
\subsection{Datasets}
We evaluate on two contrasting datasets. The \textbf{DevOps} knowledge base represents structured, domain-specific retrieval with terminology alignment and contextual grounding in a constrained space. Its composition is summarized in Table~\ref{tab:kb}, spanning six documentation types with distinct lengths and roles---from short Standard references to longer Architecture and Runbook documents. \textbf{MuSiQue} \cite{trivedi2022musique} represents open-domain compositional reasoning, where evidence must be integrated across multiple independent documents. Together they probe system behavior under both structured and multi-hop conditions.

\begin{table}[t]
\caption{DevOps Knowledge Base Composition}
\label{tab:kb}
\centering
\renewcommand{\arraystretch}{1.15}
\setlength{\tabcolsep}{5pt}
\begin{tabular}{@{}lccl@{}}
\toprule
\textbf{Document Type} & \textbf{Count} & \textbf{Avg.\ Words} & \textbf{Key Metadata} \\
\midrule
Standard     & 5  & 150 & Doc ID, Doc Type \\
Architecture & 10 & 200 & Doc ID, Doc Type \\
Runbooks     & 20 & 180 & Doc ID, Doc Type \\
Incidents    & 15 & 100 & Doc ID, Doc Type \\
Postmortem   & 15 & 120 & Doc ID, Doc Type \\
Onboarding   & 15 & 120 & Doc ID, Doc Type \\
\bottomrule
\end{tabular}
\end{table}

\subsection{Evaluation Metrics}
Table~\ref{tab:metrics} summarizes the metrics. Overall Score is an aggregate quality measure; Critical Source Recall captures whether key supporting documents are retrieved; MRR and Success@5 measure ranking quality; Citation Accuracy measures whether cited sources support the claims; Topic Coverage measures breadth; and Latency measures end-to-end response time.

\begin{table}[t]
\caption{Evaluation Metrics Used in the Study}
\label{tab:metrics}
\centering
\renewcommand{\arraystretch}{1.2}
\begin{tabular}{@{}ll@{}}
\toprule
\textbf{Metric} & \textbf{Description} \\
\midrule
Overall Score          & Aggregate performance measure \\
Critical Source Recall & Retrieval of key supporting documents \\
MRR                    & Ranking quality of retrieved results \\
Success@5              & Retrieval success within top 5 results \\
Citation Accuracy      & Correctness of cited sources \\
Topic Coverage         & Breadth of information covered \\
Latency                & End-to-end response time \\
\bottomrule
\end{tabular}
\end{table}

% =====================================================
\section{Results and Discussion}
% =====================================================
\subsection{Impact of Query Decomposition}
Decomposition behaves very differently across the two datasets. On the DevOps corpus (Table~\ref{tab:decomp}), it improves nearly every retrieval metric: overall score rises from 0.814 to 0.855, MRR from 0.556 to 0.722, Success@5 reaches a perfect 1.0, and citation accuracy climbs to 0.917. The cost is latency, which more than doubles from 21\,s to 48\,s.

On MuSiQue (Table~\ref{tab:decomp}), the picture inverts. Although citation accuracy reaches 1.0 and topic coverage rises sharply (0.609 to 0.859), ranking-based metrics collapse: MRR falls from 0.469 to 0.102 and Success@5 from 1.0 to 0.063. Decomposition therefore broadens coverage but fragments the contextual signal that ranking depends on, harming precision in genuinely multi-hop settings.

\begin{table}[t]
\caption{Impact of Query Decomposition (Baseline vs. Decomposed)}
\label{tab:decomp}
\centering
\renewcommand{\arraystretch}{1.15}
\setlength{\tabcolsep}{4pt}
\begin{tabular}{@{}lcccc@{}}
\toprule
 & \multicolumn{2}{c}{\textbf{DevOps}} & \multicolumn{2}{c}{\textbf{MuSiQue}} \\
\cmidrule(lr){2-3}\cmidrule(lr){4-5}
\textbf{Metric} & Base & Decomp. & Base & Decomp. \\
\midrule
Overall Score          & 0.814 & 0.855 & 0.786 & 0.809 \\
Critical Source Recall & 0.750 & 0.806 & 0.833 & 0.604 \\
MRR                    & 0.556 & 0.722 & 0.469 & 0.102 \\
Success@5              & 0.833 & 1.000 & 1.000 & 0.063 \\
Citation Accuracy      & 0.750 & 0.917 & 0.906 & 1.000 \\
Topic Coverage         & 0.625 & 0.558 & 0.609 & 0.859 \\
Avg.\ Latency (s)      & 21.00 & 47.66 & 21.84 & 74.87 \\
\bottomrule
\end{tabular}
\end{table}

% ---------- Placeholder: decomposition figure ----------
\begin{figure}[t]
\centering
\includegraphics[width=\columnwidth]{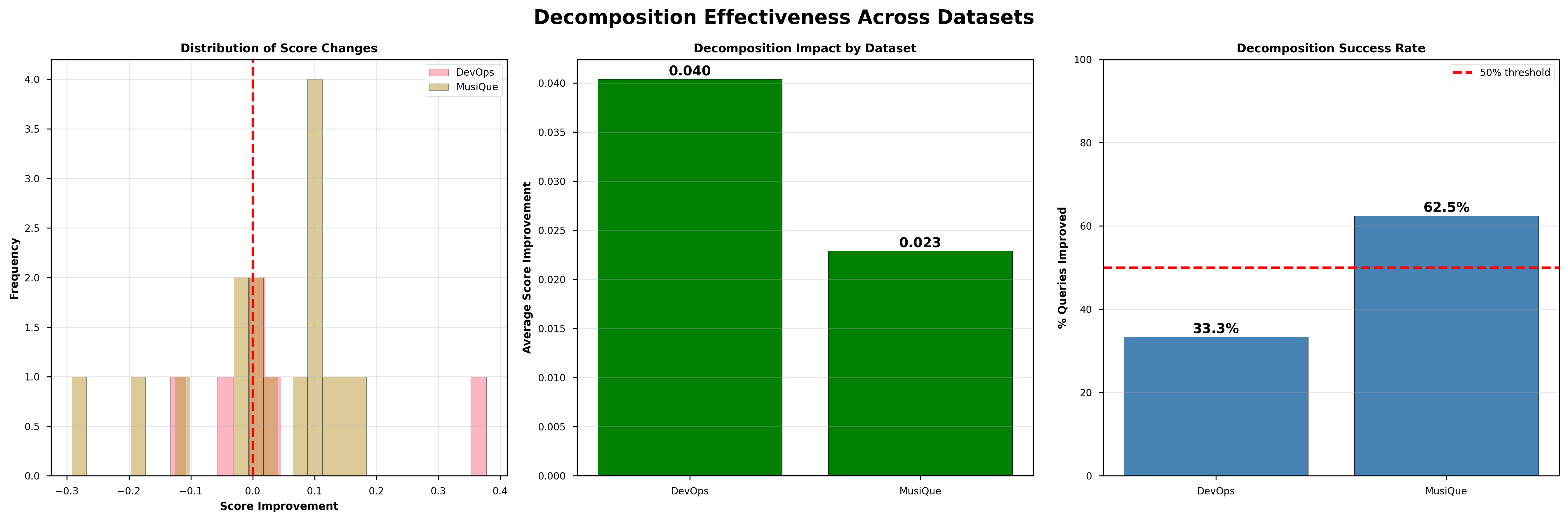}
\caption{Decomposition effectiveness across datasets.}
\label{fig:decomp}
\end{figure}

\subsection{Impact of Reflection}
The reflection mechanism yields, at best, mixed results (Table~\ref{tab:reflect}). On DevOps, overall score actually \emph{decreases} from 0.870 to 0.781 with no gain in recall or ranking, while latency roughly doubles. On MuSiQue, overall score also falls (0.721 to 0.666) and recall and MRR both decline; citation accuracy improves slightly, but latency increases roughly sixfold, from 17\,s to 104\,s. These results indicate that, as currently designed, reflection imposes a large computational cost without reliable quality gains.

\begin{table}[t]
\caption{Reflection Impact (Baseline vs.\ Full Agentic)}
\label{tab:reflect}
\centering
\renewcommand{\arraystretch}{1.15}
\setlength{\tabcolsep}{4pt}
\begin{tabular}{@{}lcccc@{}}
\toprule
 & \multicolumn{2}{c}{\textbf{DevOps}} & \multicolumn{2}{c}{\textbf{MuSiQue}} \\
\cmidrule(lr){2-3}\cmidrule(lr){4-5}
\textbf{Metric} & Base & Full & Base & Full \\
\midrule
Overall Score          & 0.870 & 0.781 & 0.721 & 0.666 \\
Critical Source Recall & 1.000 & 1.000 & 0.706 & 0.618 \\
MRR                    & 0.778 & 0.778 & 0.757 & 0.659 \\
Citation Accuracy      & 1.000 & 1.000 & 0.882 & 0.941 \\
Avg.\ Latency (s)      & 10.57 & 21.78 & 17.21 & 103.53 \\
\bottomrule
\end{tabular}
\end{table}

\begin{figure}[t]
\centering
\includegraphics[width=\columnwidth]{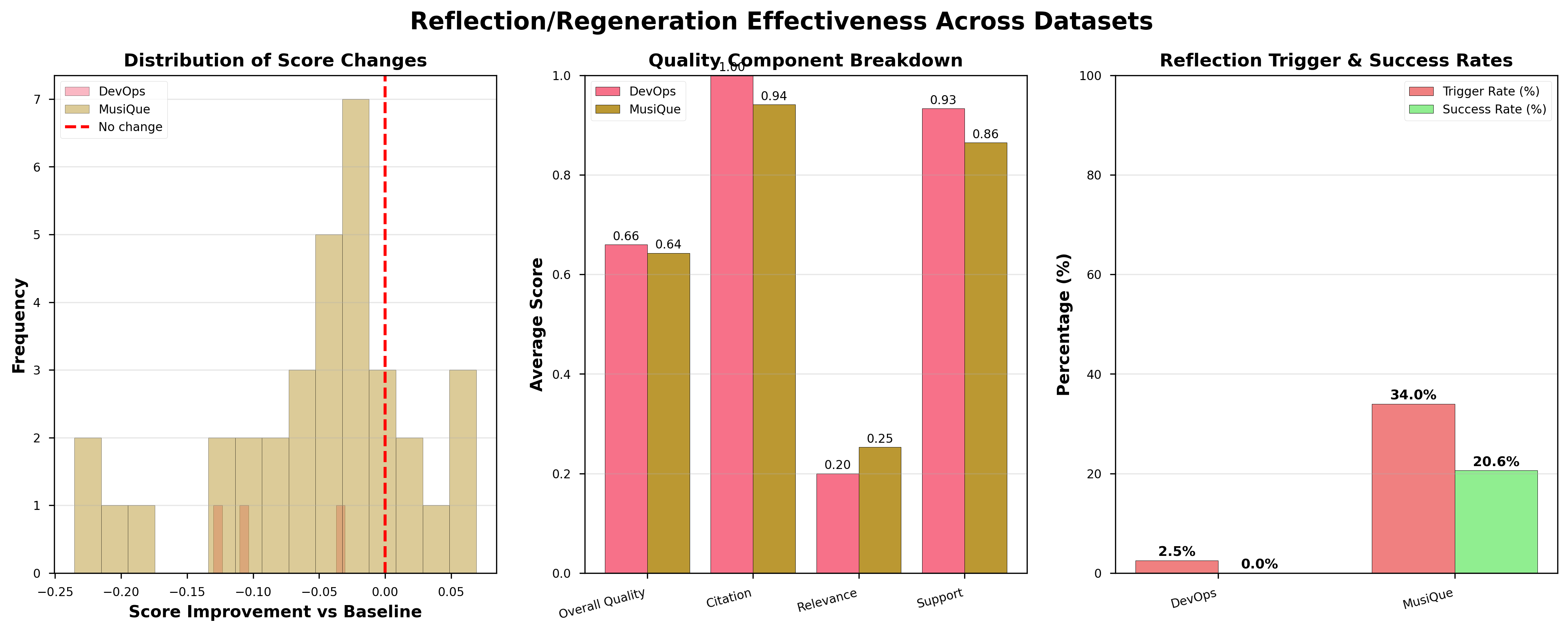}
\caption{Reflection effectiveness: quality-component breakdown.}
\label{fig:reflect}
\end{figure}

\subsection{Strategy Distribution}
The Orchestrator selects among three strategies: Standard RAG (direct retrieval), Complex RAG (multi-step retrieval without explicit decomposition), and Decomposed RAG. On DevOps, the system routes most queries through standard retrieval (69.2\%), with decomposition adopted for only about 5\% of queries---consistent with the structured, largely single-hop nature of the domain. On MuSiQue, complex queries dominate, and decomposition is adopted for roughly 16\% of queries, reducing reliance on undifferentiated complex retrieval. Notably, the distribution is identical between the decomposition-enabled and full-agentic configurations, confirming that reflection operates as a post-hoc corrective step rather than a routing mechanism.

Several patterns emerge across the two datasets. First, strategy selection is strongly dependent on dataset complexity: the structured DevOps corpus favors standard retrieval, whereas the multi-hop MuSiQue benchmark draws on advanced strategies far more often. Second, decomposition introduces a new processing pathway but never dominates the pipeline, indicating that the orchestrator invokes it selectively rather than reflexively. Third, the comparatively small share of decomposed queries---even on MuSiQue, where 49\% of queries remain in the complex-RAG category---suggests that the decomposition trigger is conservative and that not all complex queries that could benefit are actually routed for decomposition. This points directly to triggering policy as a high-value target for future optimization.

\begin{figure}[t]
\centering
\includegraphics[width=\columnwidth]{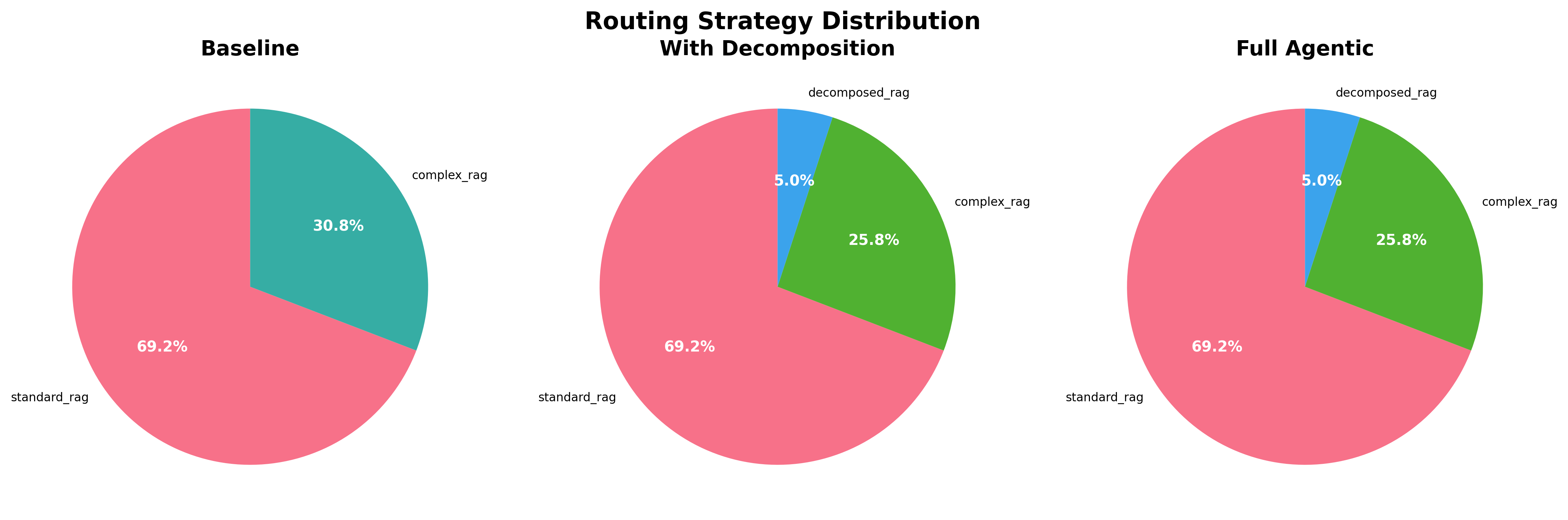}
\caption{Routing strategy distribution across DevOps Dataset.}
\label{fig:strategy-devops}
\end{figure}

\begin{figure}[t]
\centering
\includegraphics[width=\columnwidth]{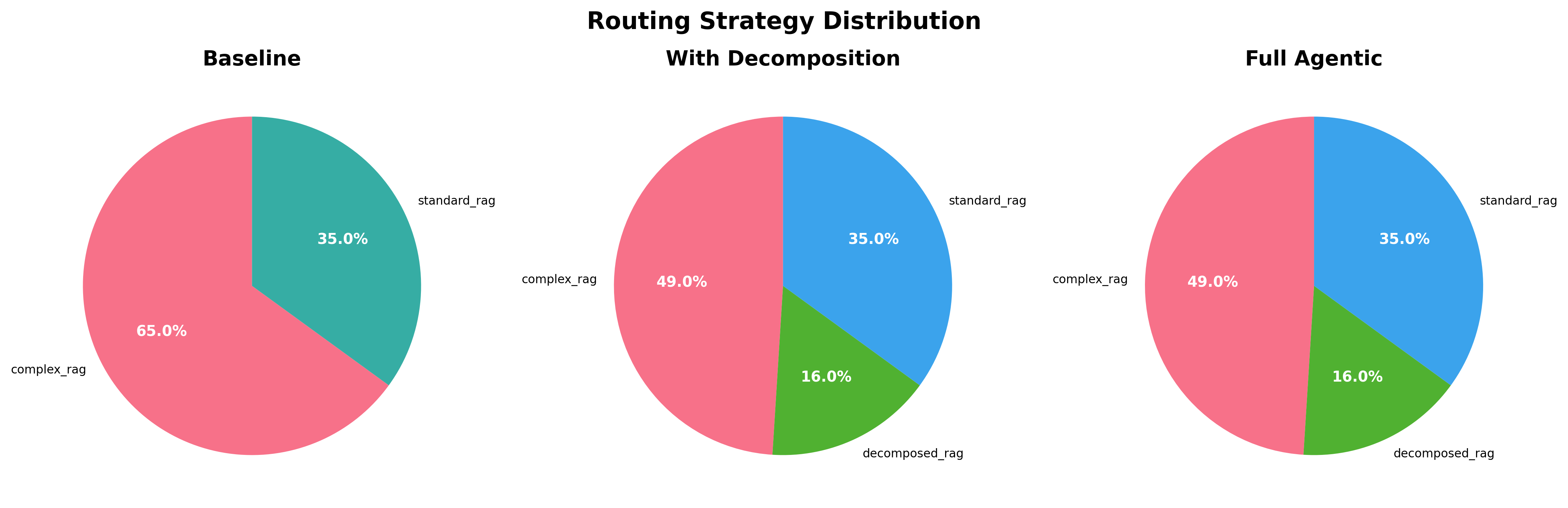}
\caption{Routing strategy distribution across MusiQue Dataset.}
\label{fig:strategy-musique}
\end{figure}

\subsection{Tradeoff Analysis}
Three tradeoffs emerge consistently. \textbf{Accuracy vs.\ latency}: every agentic addition increases response time, with reflection on MuSiQue approaching a sixfold slowdown---likely unacceptable in real-time DevOps use. \textbf{Precision vs.\ coverage}: decomposition broadens retrieval coverage but can dilute the contextual signal that ranking metrics depend on, lowering MRR and Success@5 in multi-hop settings. \textbf{Control vs.\ efficiency}: modular agents improve interpretability and enable adaptive behavior, but at the cost of additional LLM calls and retrieval operations that limit scalability.

\subsection{Summary of Findings}
Decomposition helps in structured domains where chained reasoning maps cleanly onto well-defined relationships, but it is not universally beneficial and degrades ranking precision on harder multi-hop data. Reflection offers limited, inconsistent gains relative to its cost. And because a large share of queries are handled well by standard RAG, adaptive orchestration---applying expensive strategies only when warranted---is essential rather than optional.

\subsection{Limitations and Threats to Validity}
Several limitations qualify these findings and frame the scope of our claims. First, the DevOps knowledge base is modest in scale (80 documents, roughly 10{,}000 words) and was generated with LLM assistance before manual validation; while it is designed to emulate realistic structural heterogeneity, results on a larger production corpus may differ, and the absolute metric values should be read as indicative rather than definitive. Second, the evaluation query sets are correspondingly small, so per-metric differences---particularly the sharp ranking drops on MuSiQue---carry wide confidence intervals and would benefit from larger samples and significance testing. Third, the entire study uses a single generator (Llama-3.1-8B-Instruct) and a single embedding model; the interaction between agentic strategies and model capacity is unexplored, and a stronger generator might absorb some errors that decomposition currently compensates for. Fourth, the orchestrator's routing relies on rule-based heuristics, so the observed strategy distributions reflect those specific rules rather than an optimal policy. Finally, latency figures are environment-dependent and were measured under local quantized inference; they capture relative trends across configurations rather than portable absolute timings. We view these constraints not as undermining the central result---that agentic benefits are domain-dependent and cost-laden---but as delimiting the conditions under which it was established.

% =====================================================
\section{Conclusion and Future Work}
% =====================================================
We presented an agent-orchestrated adaptive RAG system that combines query classification, decomposition, answer evaluation, and a bounded reflection loop to select retrieval strategies dynamically. Across a structured DevOps corpus and the multi-hop MuSiQue benchmark, query decomposition improved retrieval quality in the structured domain but reduced ranking precision in the multi-hop setting, and reflection delivered marginal quality gains at substantial latency cost. The clear lesson is that agentic enhancements must be applied selectively and with explicit cost awareness rather than uniformly.

\subsection{Adaptive Strategy Selection}
A central limitation we observed is that decomposition and reflection are triggered by static, heuristic rules. Future work can replace these with learning-based strategy selection, in which the system predicts the optimal reasoning strategy from query features. Reinforcement learning or meta-learning could be used to balance quality against efficiency directly, learning when decomposition genuinely helps---as in the DevOps domain---and when it harms ranking, as on MuSiQue.

\subsection{Cost-Aware Optimization}
Given the large latency penalties measured for agentic components, orchestration should become explicitly cost-aware. Incorporating a cost model that accounts for retrieval time, the number of reasoning steps, and per-call inference cost would let the orchestrator weigh expected quality gains against their computational price, which is essential for latency-sensitive DevOps deployments where the sixfold reflection slowdown we observed would be unacceptable.

\subsection{Improved Query Decomposition}
The current decomposition strategy can be refined to address the precision loss seen on multi-hop data. Promising directions include generating more semantically coherent sub-queries, explicitly preserving contextual dependencies between sub-tasks so that ranking signal is not fragmented, and integrating structured reasoning representations such as knowledge or reasoning graphs to capture inter-step relationships that flat sub-query lists miss.

\subsection{Enhanced Reflection Mechanisms}
Reflection in this work is a bounded retry loop. More sophisticated designs could introduce fine-grained error detection and classification, selective re-execution of only the pipeline components implicated in a failure rather than the whole query, and confidence-based stopping criteria that halt refinement once marginal gains fall below a threshold. Such targeted reflection would aim to recover its quality benefits while shedding much of its current cost.

\subsection{Integration with External Tools}
The framework can be extended beyond retrieval and generation to incorporate additional tools---knowledge graphs, live APIs, or symbolic reasoning engines---following the tool-use paradigm of ReAct \cite{yao2022react} and Toolformer \cite{schick2023toolformer}. In a DevOps context this could mean querying live monitoring systems or ticketing APIs, enabling responses grounded in the current state of a system rather than static documentation alone.

\subsection{Scalability and Real-World Deployment}
Finally, the system should be evaluated at larger scale and under realistic conditions, including noisy queries, much larger corpora, and concurrent user load. Establishing robustness, throughput, and operational cost in production settings is necessary before agentic RAG can move from a research prototype to a deployed assistant. Recent advances in adaptive retrieval and cost-aware agentic reasoning reinforce the importance of these directions \cite{maharjan2026chunking, mishra2026sok}.

\bibliographystyle{IEEEtran}
\bibliography{references}

\end{document}